 \documentclass[final,5p,times,twocolumn]{elsarticle}
\setcitestyle{number}
\usepackage[utf8]{inputenc}
\usepackage{bm,amsmath,amssymb}
\usepackage{graphicx}
\usepackage{subfigure}
\usepackage{color}
\usepackage{hyperref}
\usepackage{multirow}
\usepackage{soul}
\usepackage{comment}

\let\oldalign\align
\def\align{\linenomath\oldalign}

\newcommand{\Dov}{D_\text{ov}}
\newcommand{\pp}{$pp$}
\newcommand{\ep}{$ep$}
\newcommand{\sqrtS}{$\sqrt{s}$}
\newcommand{\JetPt}{$p_{\rm T,jet}$}
\newcommand{\Delphi}{$\Delta\phi$}
\newcommand{\phiLead}{$\phi^{\mathrm{L}}$}
\newcommand{\phiSubLead}{$\phi^{\mathrm{SL}}$}

\newcommand{\pTjetL}{$p_{\rm T, jet}^{\rm L}$}
\newcommand{\pTjetSubL}{$p_{\rm T,jet}^{\rm SL}$}

\newcommand{\AveCosTwoDPhi}{$\langle \cos (2\Delta\phi) \rangle$}

\begin{document}

\title{ Probing QCD instantons using jet correlation observables in proton-proton collisions at the LHC  }

\author[first,second]{Sayak Guin\corref{cor1}}

\author[first,second]{Swagatam Tah\corref{cor1}}

\author[third]{Nihar Ranjan Sahoo}

\author[first,second]{Sayantan Sharma}
\affiliation[first]{The Institute of Mathematical Sciences, Taramani, Chennai 600113, India}
\affiliation[second]{Homi Bhabha
National Institute, Training School Complex, Anushakti Nagar, Mumbai 400094, India}
\affiliation[third]{Department of Physics, Indian Institute of Science Education and Research (IISER), 
Tirupati, Andhra Pradesh, 517619, India}

\cortext[cor1]{Corresponding authors}

\date{\today}
\begin{abstract}

Discovery of instantons in colliders will provide experimental evidence for the topological properties of the QCD vacuum. In this work, we propose jet correlation observables that can unambiguously discriminate between instanton-induced processes and perturbative hard scattering events in $pp$ collisions at the LHC for a specific range of center-of-mass energies of the produced hadrons. By calculating the instanton sizes and their separations in 2+1 flavor QCD with physical quark masses, we provide a qualitative understanding of why the semiclassical methods reliably predict the cross-section of instanton-induced processes at the particular center-of-mass energies we are focusing on. Our proposal is directly applicable to future $ep$ measurements at the Electron-Ion Collider, offering a cleaner environment to probe instanton-induced processes.    
\end{abstract}
\maketitle


\section{Introduction}

The theory of strong interactions, Quantum Chromodynamics (QCD), is inherently non-perturbative, 
leading to color confinement and chiral symmetry breaking~\cite{Gross:2022hyw}. These phenomena are 
driven by topological properties of the QCD vacuum~\cite{Schafer:1996wv}, which is infinitely 
degenerate. Each degenerate ground-state is labeled by an integer $n$.  Tunneling between two neighboring 
vacua $\lvert{n} \rangle \to \lvert n\pm 1 \rangle$ is mediated by gluon field configurations known
as (anti)instantons, which carry a topological charge $Q=\pm 1$, respectively.  Instantons are thus zero 
energy solutions of the QCD Hamiltonian. These are also classical solutions with finite action in four dimensional 
Euclidean spacetime~\cite{Belavin:1975fg,tHooft:1976snw} which are characterized by a single scale which is 
their size. Moreover, instantons induce local violation of the charge-parity (CP) symmetry, even though the 
QCD vacuum respects CP globally. Substantial evidence from phenomenological models~\cite{Schafer:1996wv} and non-perturbative lattice studies~\cite{DeGrand:2000gq} suggests that instantons play a role in chiral symmetry breaking and 
in hadron mass generation. However, direct observation of instanton-induced events has so far remained elusive and thus one of the most coveted goals of collider experiments. \\

Specific signatures of QCD instanton-mediated processes in a deep-inelastic scattering (DIS) event involve final 
states with a large multiplicity of hadrons, including strange hadrons, emerging from a \emph{fireball}-like topology 
in the instanton rest frame with a large total transverse energy~\cite{Ringwald:1994kr}. Such hard events  are forbidden in usual perturbative scattering processes in QCD but their rates induced by instantons may be substantial and has 
been estimated using instanton perturbation theory~\cite{Moch:1996bs, Ringwald:1998ek}. The H1 Collaboration reported 
a dedicated search for instanton-induced events at HERA in a phase-space region where the usual DIS background is 
reduced to the percent level by standard Monte-Carlo estimates~\cite{H1:2002lra}; however, no conclusive signal has 
been observed~\cite{H1:2016jnv,ZEUS:2003qkm}.

There are two important aspects, one theoretical and the other experimental, which needs to be understood 
in order to address why instanton-induced events could not be unambiguously identified in DIS experiments.
First, the event generators which can simulate instanton-induced events rely on the valley-method~\cite{Balitsky:1986qn} 
to calculate the relevant cross-sections~\cite{Ringwald:1999jb}. Such a method relies on the fact 
that the typical separation between instantons and anti-instantons is larger than their size. Fiducial cuts 
for instanton-induced events have been estimated using lattice inputs on instanton size distributions and 
separations~\cite{Ringwald:1999ze} but only in quenched QCD. It remains to be investigated for what typical size 
of instantons in QCD with physical quarks, the instanton perturbation theory remains a valid 
description. Secondly, it is experimentally challenging to disentangle the instanton-induced events from the 
overwhelmingly large background due to traditional DIS and also perturbative QCD processes. 

The typical analysis strategy involves identifying observables, which in this case are the transverse energy 
($\lesssim$ a few GeV) and sphericity of a dense population of hadrons in a narrow band in pseudo-rapidity 
and the reconstructed quark virtuality~\cite{H1:2002lra} to discriminate signal from the
dominant background. However, a proper implementation of the required fiducial cuts and the inefficiency of the usual 
DIS event generators in a region of phase space where instanton events are predicted, motivates for exploring more 
cleaner experimental set-up. The search for the instanton-induced processes thus could be particularly well suited 
in  \pp\ collisions~\cite{Khoze:2020tpp,Amoroso:2020zrz,Khoze:2021jkd,Tasevsky:2022sch} at the LHC energies 
or \ep\ collisions at the EIC.  These systems allow us to explore the low $x$-regime ($\lesssim 10^{-3})$ 
where instantons are copiously present. For an instanton-induced process, the center-of-mass energy 
$\sqrt{s^{'}}$ of the produced hadrons~\cite{Amoroso:2020zrz} sets a bound on maximum size of instantons 
participating in the process.  
A high multiplicity of low energy hadron jet with a total $\sqrt{s^{'}}> 50$ GeV and small transverse momentum 
$p_T \gtrsim 10$ GeV is a typical example of an instanton-induced process which is distinct from events with 
perturbatively produced dijets with $p_T> 25$ GeV and soft non-perturbative inelastic multi-parton interactions (MPI).

 In this Letter, we propose a novel approach based on jet correlation observables, in particular jet acoplanarity as a 
 probe of instanton-induced processes in proton–proton collisions. The proposal is based on the premise that instanton 
 processes are expected to produce a relatively isotropic distribution of hadrons, in contrast to the back-to-back 
 topology characteristic of perturbative dijet events. This difference can be quantified using the azimuthal separation 
 \Delphi = \phiLead - \phiSubLead, between a leading and a subleading jet,  and its harmonic moments. By constructing 
 these observables within a jet-based event generator, we exploit the sensitivity of modern jet reconstruction 
 techniques to both hard and soft radiation patterns. We demonstrate that, while perturbative QCD processes exhibit 
 strong angular correlations, instanton-like event topologies lead to a suppression of such correlations, providing a 
 potential discriminant. The instanton-induced cross-sections calculated in the event generators crucially rely on the 
 validity of semiclassical approximation at the particular centre-of-mass energy of the produced hadrons.  
 By calculating, for the first time, the distribution of (anti)-instantons sizes and their separations in 2+1 flavor 
 QCD with physical quark mass using ab-initio lattice techniques, we provide a qualitative understanding why 
 semiclassical approximation might be reliable for estimating instanton-induced scattering cross-sections in the 
 kinematic range we are suggesting. The lattice results presented in this work thus provide insights into the 
 instanton-antiinstanton distribution in the QCD vacuum. Which among them will participate in a scattering event is 
 dynamically determined based on the energy scale.

This Letter is organized as follows: Section~\ref{Sec:LatticeQCD} describes in detail the techniques used to 
determine the distribution of instanton sizes and their relative separation using lattice QCD. The simulation framework, 
the proposed observables in \pp\ collisions, and  the key results and their implications are presented in 
Sec.~\ref{Sec:SimInstanton}. Finally, Sec.~\ref{Sec:summary} summarizes the main findings and outlines future 
prospects.

\section{Constraining the instanton vacuum properties from Lattice QCD}
\label{Sec:LatticeQCD}

\subsection{Methodology}
Using first principles lattice techniques, we estimate the size and distance distribution among the 
instantons present in the QCD vacuum. By virtue of the Atiyah-Singer index theorem~\cite{Atiyah:1963zz}, 
the (anti)instantons in QCD are related to the (left)right-handed zero eigenmodes of the massless Dirac 
operator. Hence, one can identify the location and shape of the instantons by studying the properties of 
the Dirac zero eigenmodes. It is thus crucial for our study to use fermion Dirac operator which has an 
exact chiral symmetry on the lattice and an exact index for our study. We use overlap fermion discretization 
$\Dov=1+\gamma_5~\text{sgn}(\gamma_5 D_W)$ since this is the only lattice Dirac operator~
\cite{Narayanan:1994gw, Neuberger:1998my} that satisfies both of these 
criteria~\cite{Ginsparg:1981bj,Luscher:1998pqa,Hasenfratz:1998ri}. Overlap 
fermions are computationally challenging to implement, hence we perform several numerical optimizations. 
First, the matrix sign function that goes into the definition of the overlap operator was implemented 
in terms of the first $30$ Dirac eigenvectors of the Hermitian Wilson operator $\gamma_5 D_W$ exactly 
and as a Zolotarev rational polynomial consisting of $25$ terms in rest of the subspace. The matrix 
sign function was implemented with a precision of $\lesssim 10^{-9}$ on the gauge ensembles which were 
generated~\cite{Gavai:2024mcj} with the M\"{o}bius domain wall fermions (MDWF) discretization~\cite{Brower:2004xi} 
for quarks and Iwasaki gauge action. The choice of MDWF for generating QCD gauge ensembles was motivated 
from the fact that these have almost exact chiral symmetry and are closely related to the overlap 
Dirac operator, though computationally less prohibitive. We want to remind here that the overlap Dirac 
operator was used as a probe to detect the density and size distribution of the instantons 
in the QCD configurations generated with MDWF.  Since the overlap Dirac operator satisfied the Ginsparg-Wilson 
relation with a precision $\lesssim 10^{-9}$, its zero eigenmodes always had a chirality $\pm 1$ just like in the 
continuum, clearly distinguishable from the near-zero eigenmodes whose chiralities are finite and non-zero. This 
allowed for a clear identification of the instantons (and anti-instantons) in the QCD configurations. The eigenvalues 
of the overlap Dirac operator was calculated with an optimized version of the Kalkreuter-Simma ritz 
algorithm~\cite{Kalkreuter:1995mm} with restarts. 

The lattice box chosen in our simulations has $N_s=32$ sites along each spatial direction and a 
temporal extent $N_\tau=8$. In physical units the box size is $\sim 5.3$ fm, which is large 
enough to ensure physical reliability of our results. The QCD ensembles were generated at a temperature 
$T = 149$ MeV, in the hadronic phase of QCD. At this temperature, topological fluctuations of QCD vacuum 
quantified in terms of topological susceptibility, agrees well with the chiral perturbation theory prediction 
that describes the QCD vacuum fairly well at $T = 0$~\cite{Bonati:2015vqz}. In fact the non-perturbative 
magnetic scale that characterizes QCD vacuum, denoted by the spatial string tension, is similar~
\cite{Bala:2025ilf} to the string tension that describes the confining potential in QCD at $T = 0$.
The light and strange quark masses are tuned such that the extracted pion and kaon mass are physical, 
i.e. $135$ and $435$ MeV, respectively. In order to compare our results with continuum QCD, we convert 
the lattice spacing $a$ in physical units in terms of the mass of the $\Omega$-baryon using 
$am_\Omega=1.53$~\cite{Bhattacharya:2014ara} calculated at a slightly larger spacing, however the 
dependence on the lattice cut-off is very mild for this observable. We have analyzed $84$ independent 
gauge configurations in our study in order to have a reasonably good control over statistical errors. 
We next discuss our results from lattice investigations of instanton size and their relative distances.

\subsection{Size distribution}

\begin{figure}[!h]
    \centering
    \includegraphics[width=0.48\textwidth]{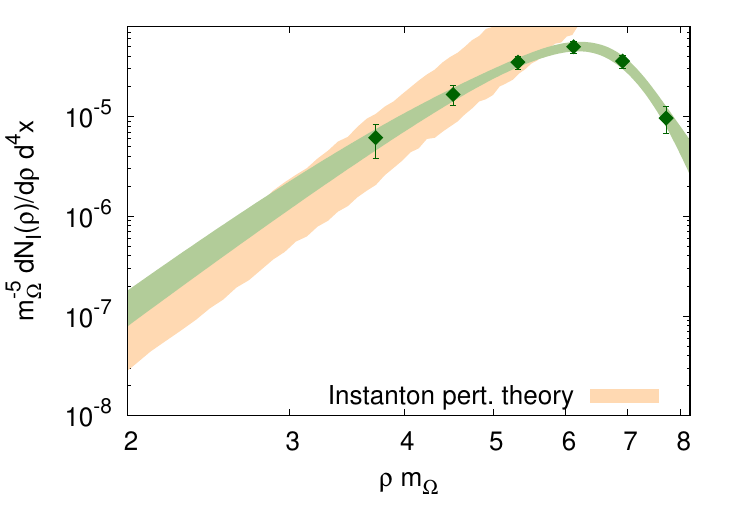}
    \caption{Size distribution for instantons in 2+1 flavor QCD calculated on the lattice are represented as points whose functional dependence (green band) is compared with the prediction from instanton perturbation theory for $N_f=3$ (yellow band).}
    \label{fig:radiusdistributionpervsnonpert}
\end{figure}

In order to calculate the cross section for scattering off the (anti-)instantons, one needs information 
about their size distribution and their relative distances from each other. The instanton localized at a 
position in 4D Euclidean spacetime, $x_0\equiv (\mathbf{x}_0,\tau_0)$ is characterized by size parameter 
$\rho$ which is the only scale in the problem. A zero eigenmode of the Dirac operator $\psi_0(x)$ calculated 
in presence of an isolated instanton is centered at the same position as that of the instanton, whose 
probability density $\rho_d(x)= \psi_0^\dagger(x)\psi_0(x)$ in the continuum \cite{Schafer:1996wv} is 
denoted as 
\begin{equation}\label{Fit 4D profile}
    \rho_d(x)=\frac{2\rho^2}{\pi^2}\frac{1}{\Big((x-x_0)^2+\rho^2\Big)^{3}}.
\end{equation}
The maximum of this density is situated at $x_0$ such that  
$\rho_d(x_0)=\rho^\text{max}_d= 2/(\pi^2 \rho^4)$. 
Using this fact we have extracted the size parameter $\rho$ from a fit of the zero mode densities calculated 
on the lattice with the expression for $\rho^\text{max}_d$ in the continuum. We then performed a binning in terms 
of $\rho$ in bin sizes of $\sim0.52 a$. The resultant distribution $1/m_\Omega^5.dN_{I}(\rho)/d\rho. d^4x$  
of the size per unit volume appropriately normalized by $m_\Omega$ thus extracted in our study is shown in 
Fig.~\ref{fig:radiusdistributionpervsnonpert}. From the plot it is evident that the size distribution of instantons 
in QCD grows with increasing size until it attains a maximum at $\sim 0.65$ fm beyond which the distribution 
function drops rapidly. We have performed a fit to our lattice data with an ansatz 
$(A\tilde{\rho})^7~\rm{e}^{-(c\tilde{\rho})^2}/(1+(b\tilde{\rho})^d)$~\cite{Ringwald:1999ze} 
where $\tilde{\rho}\equiv\rho m_\Omega$, which resulted in $A=0.053(4)$,  $b=0.148(7)$, $c=0.21(5)$ 
and $d=18.4(3.9)$ determined with a $\chi^2/\text{d.o.f}=1.4$. The large sized instantons are exponentially 
suppressed as $\rho^2$, which is a result of the non-perturbative nature of QCD interactions at large distance 
scales. In the dilute gas approximation, the size distribution for QCD instantons with $N_f$ flavors of light 
quarks with mass $m_i,~i=1,2,..,N_f$  can be calculated analytically at leading order in strong coupling, 
$\alpha_s\equiv\alpha_s^{\overline{MS}}(\bar{\mu})$ in the ${\overline{MS}}$ scheme at a scale $\bar{\mu}$ 
which results in 
\begin{eqnarray}
\nonumber
\frac{dN_{I}(\rho)}{d\rho~d^4x} = D(\rho) \left(\prod_{i=1}^{N_f}\rho m_i(\bar{\mu})\right) \left(\rho\bar{\mu}\right)^{N_f\gamma_0\alpha_s/4\pi} \\ \nonumber
    \text{where,}~D(\rho)= \frac{d_{\overline{MS}}}{\rho^5} \left(\frac{2\pi}{\alpha_s}\right)^6 \exp\left(-\frac{2\pi}{\alpha_s}\right) \left(\rho\bar{\mu}\right)^{\beta_0+(\beta_1-12\beta_0)\alpha_s/4\pi}.
\end{eqnarray}
Here $\gamma_0=3(N_c^2-1)/N_c$, where $N_c=3$ and $\beta_0,~ \beta_1$ are the first and second coefficients of 
the QCD beta function with $d_{\overline{MS}}=0.00250~e^{0.291746N_f}$~\cite{Ringwald:1999ze}. We have compared our 
best-fit parametrization of the lattice data shown as a green band in Fig~.\ref{fig:radiusdistributionpervsnonpert} 
with the size distribution in QCD with $N_f=3$ massless quark flavors calculated within the dilute gas approximation, 
shown in the same plot as a yellow band. The perturbative band is obtained as a result of the variation of the 
renormalization scale $\bar{\mu}$ between $1$-$100$ GeV. It is clearly evident that the semiclassical calculation 
starts to describe the lattice data only for instanton sizes $\rho m_\Omega< 5$ beyond which the deviation is 
significant. In a lattice study, one cannot probe physical quantities at length-scales of the order of a lattice 
spacing $a$ or less, which in our case is $a\sim 0.16$ fm. Hence we had to extrapolate our fit to small sizes for 
a comparison. Moreover the average size of instantons obtained in our study is $\langle \rho \rangle= 0.65(1)$ fm.

\subsection{Distance distribution}

Having extracted the information about the size distribution of instantons we next calculate the typical 
distances between them. In order to do so, we recall that exact Dirac zero modes represent isolated 
instantons, whereas any interactions among them, however weak, will result in the proliferation of the 
near-zero modes in the Dirac eigenvalue spectrum. Analyzing the density distribution of the near-zero modes, 
we could identify distinct peaks due to instantons denoted as $I$ (and anti-instantons $\bar{I}$). We also measured 
the chiral densities $\rho_5(x)=\psi^\dagger(x)\gamma_5\psi(x)$ of the near-zero Dirac eigenvalues $\psi(x)$ so as 
to clearly distinguish between instantons from anti-instantons. Selecting the peaks in near-zero mode profiles 
with positive and negative chiral densities we could estimate the typical distances $R$ between two like and 
opposite charged topological objects. By appropriately binning these data in bins of size $\Delta R$ we could 
calculate the distribution of these separations per unit four-volume denoted as 
$\frac{dN_i}{d^4x~d^4 R}=\frac{\Delta N_i}{2\pi^2 R^3\Delta R d^4x}$,
where $i \in \{I\bar{I}, II, \bar{I}\bar{I}\}$ represents the possible pairs of objects whose separations are 
measured. Analyzing our data provides us with two important insights. First the average distance between $II$ 
and $\bar{I}\bar{I}$ are the same, which is true for a CP-even theory like QCD and is a check that our analysis 
has been performed correctly. Secondly, the average separation between like-charged $II$ pairs is smaller than the 
unlike-charged $I\bar{I}$ pairs hinting to the fact that presence of fermions lead to attractive interactions 
between the like-charged pairs. 

\begin{figure}[!h]
    \centering
    \includegraphics[width=0.48\textwidth]{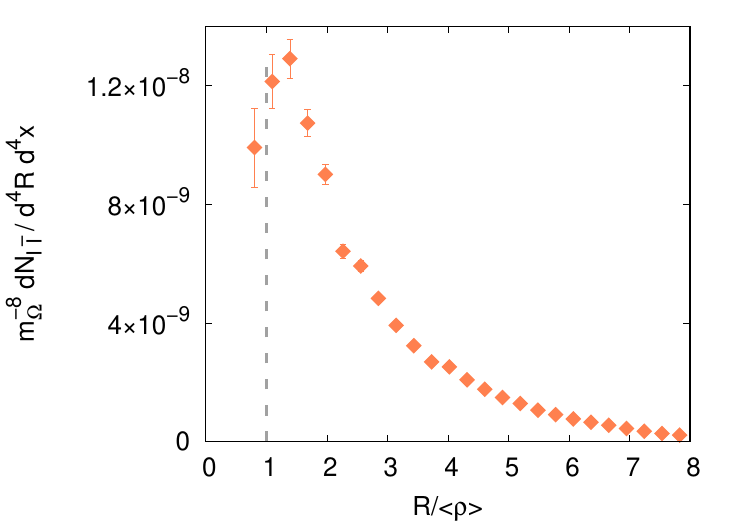}
    \caption{The distribution $\frac{1}{m_\Omega^8} \frac{dN_{I\bar{I}}}{d^4 R~ d^4x}$ of the separation $R$ between instanton and anti-instanton pairs versus $R/\langle \rho \rangle$, where the average size is $\langle \rho \rangle\sim 0.65$ fm. }
    \label{fig:DistanceDistrIbarI}
\end{figure}

The normalized distribution for the separation between instanton and anti-instanton pairs 
$\frac{1}{m_\Omega^8}\frac{dN_{I\bar{I}}}{d^4x~d^4 R}$ as a function of $R$ normalized by the 
average size of instantons, is shown in Fig.~\ref{fig:DistanceDistrIbarI}. 
The distribution has a long tail, the average distance is estimated to be 
$\langle R \rangle \sim 2.43(4)\langle \rho \rangle$. This emphasizes the fact that 
the average distance between instantons of size  $\rho \lesssim 0.5$ fm is large 
enough for the dilute gas approximation to be valid for small instantons even in the hadron 
phase of QCD. Event generators which provides hadron multiplicities calculated within instanton 
perturbation theory typically require the maximum $\sqrt{s^{'}}$ of these hadrons as an input. 
Phenomenological considerations~\cite{Khoze:2019jta} are used to estimate the typical size of instantons 
that contribute to such a process, which is related as $\sqrt{s^{'}}=(10$-$30)/\rho$. Thus for $\sqrt{s^{'}}=50$ 
GeV, instantons of maximum size $0.04$-$0.12$ fm will contribute. Our lattice calculations show that such 
instantons are indeed widely separated, hence can be described perfectly within the semiclassical instanton 
perturbation theory.

\section{Simulations of instanton-induced events}
\label{Sec:SimInstanton}


\subsection{Methodology}
\label{sec:MethodSimu}

Our motivation is to propose observables which are most sensitive to the instanton-induced processes in a 
typical hadron collision; here we consider \pp\ collisions at the LHC energies. In order to motivate for experimental 
measurements of such observables we first test their effectiveness within a typical event-generator which mimics 
the experimental conditions. \textsc{SHERPA}~\cite{Gleisberg:2008ta,Sherpa:2019gpd} is one such event generator 
which has been used extensively to simulate instanton-induced processes. A dominant process~\cite{Khoze:2020tpp} 
which is initiated by gluons representing an instanton configuration with topological charge $Q=1$ is
\begin{equation*}
    g+g\longrightarrow n_g\times g+\sum_{f=1}^{N_f}\left(q^f_{R}+\bar{q}^f_{L}\right)~
\end{equation*}
where $n_g$ is the number of gluons in the final state which varies event-to-event following a Poissonian
distribution and the quark content is fixed to $N_f=5$. The invariant mass of the final hadrons 
defines the maximum inverse size of the instantons producing them. In this work, we simulate \pp\ events at 
center of mass energy $\sqrt s=13$ TeV. All final state particles, within pseudo-rapidities $|\eta|<2$ and 
 $2\pi$ azimuthal angle, are considered for reconstructing jets discussed in~\ref{sec:Jetreco}. 
Two sets of events are generated, one for $\sqrt{s^{'}}=50$ GeV and another for $\sqrt{s^{'}}=100$ GeV with a 
subprocess cross-sections of $3.89\times 10^{4}$ pb  and $5.44\times 10^{2}$ pb, respectively, as provided 
within SHERPA. A quark-initiated instanton subprocess is also possible but is suppressed compared to the 
gluon-initiated process due to the overwhelming dominance of the gluon parton distribution function at 
small $x$, accessible at LHC energies.

The cross-section for the instanton-induced subprocess requires the instanton-anti-instanton action, 
$S_{I\bar{I}}=S_I+S_{\bar{I}}+U_\text{int}\Big(\rho,\bar\rho,R,C\Big)$, which consists of the action 
for an isolated instanton and an anti-instanton, and their mutual interactions $U_\text{int}$. The non-perturbative 
interaction potential among instantons and anti-instantons cannot be yet calculated but can be modeled in the 
semiclassical regime, where the average size of (anti)-instantons is much smaller compared to the distance $R$ 
between them. It is also a function of their relative color-orientations denoted by $C$. Depending on 
these parameters, the interaction potential can be calculated within the valley approximation which uses 
conformal invariance of the classical Yang-Mills theory. Identifying the valley and integrating over the fields 
in all other orthogonal directions and subsequently over the slow valley degrees of freedom, it turns out 
that the dominant contribution from the cross-section of an instanton subprocess sets the value of the average 
number of gluons $n_g$ in the final state, for a particular input instanton mass. In \textsc{SHERPA}, 
the parton showers in the initial and final states are initialized at a scale $\mu_Q$, the maximum among the 
transverse momentum carried by the outgoing partons. After the parton shower terminates, events are further 
supplemented with the usual multi-parton interactions leading to the hadronization of emerging partons.

Perturbative dijet production constitutes the dominant background to instanton-induced multi-parton processes 
in \pp\ collisions. Since leading and sub-leading jet correlations in the dijet event arises from higher-order QCD 
effects such as initial and final-state radiation and recoil, their detailed modeling are known to be dependent on 
the choice of the event generator. To assess the robustness of the perturbative baseline and observables, we also 
compare results from \textsc{PYTHIA8} with Detroit tune~\cite{Aguilar:2021sfa}, hereafter simply referred to as 
\textsc{PYTHIA8}. \textsc{SHERPA} and \textsc{PYTHIA8} employ different parton-shower formalisms, recoil schemes, 
and matrix-element matching strategies.

\subsection{Jet reconstruction and selections}
\label{sec:Jetreco}
Jets are reconstructed for both instanton-induced processes and dijet events simulated using \textsc{SHERPA}, whereas 
only dijet events from \textsc{PYTHIA8}. In this study, a small jet radius is essential to resolve the mini-jet–like 
structure of instanton-induced events. This avoids artificial reconstruction of dijet topology that is absent at 
the partonic level. We choose $R=0.3$ in order to provide an optimal balance between resolving the multi-parton 
structure of instanton-induced events and maintaining robustness against soft-radiation and hadronization effects. 
Although a systematic study of the jet-radius dependence could provide further insights into the interplay between 
perturbative and non-perturbative QCD effects, such an investigation is beyond the scope of our present work.

\begin{figure}[!h]
    \centering
    \includegraphics[width=0.4\textwidth]{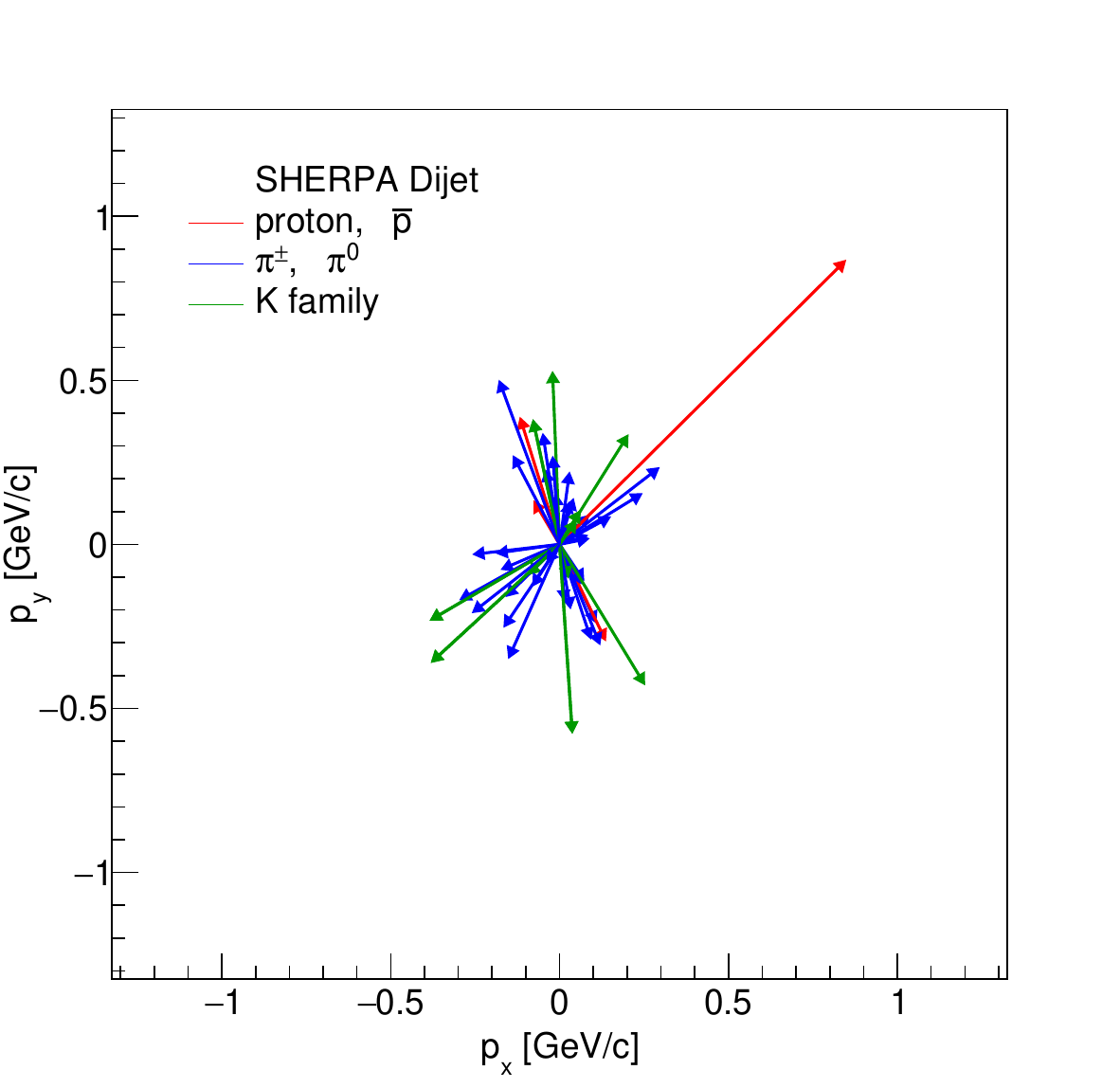}
     \includegraphics[width=0.4\textwidth]{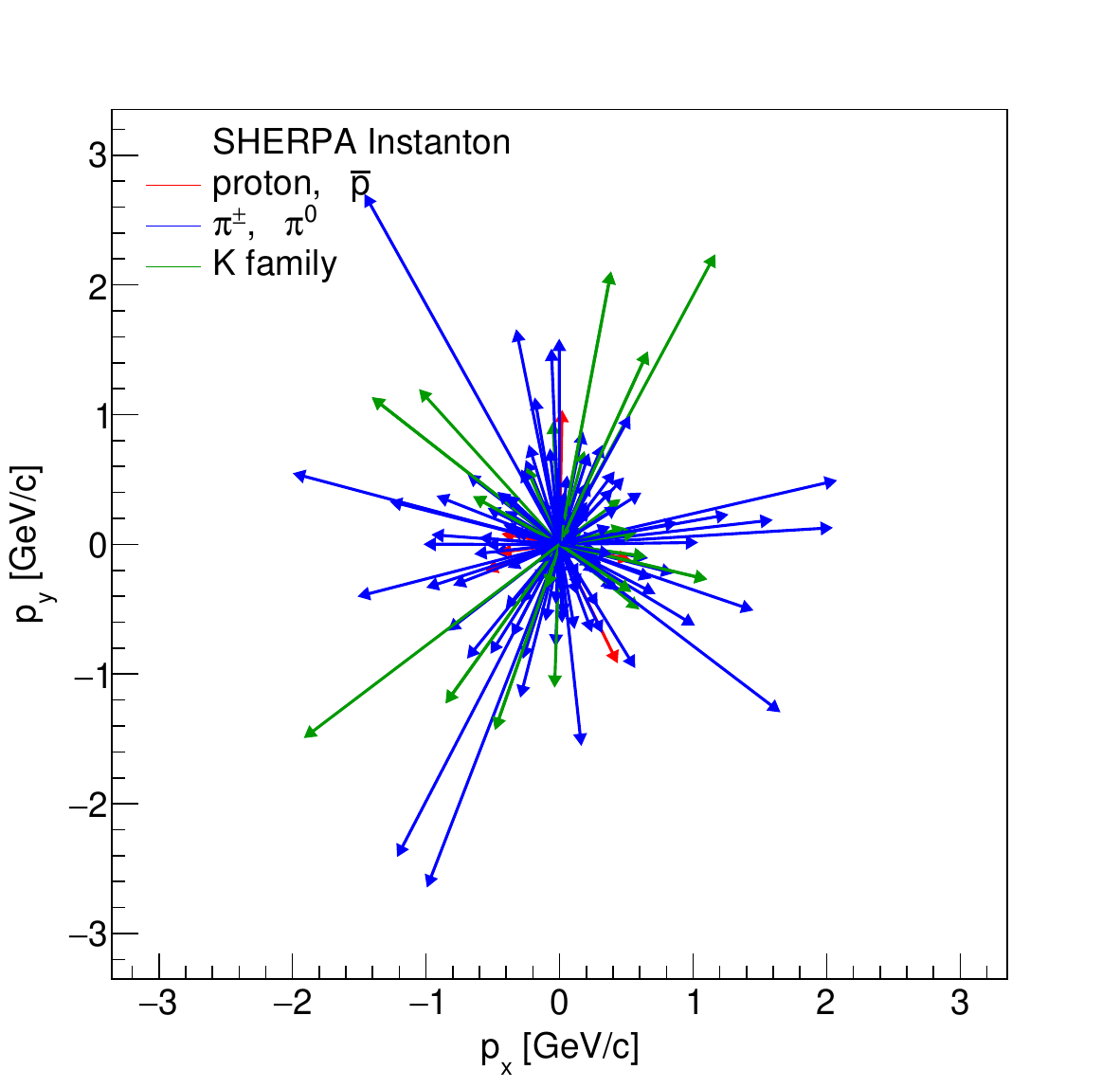}
    \caption{A dijet (left) and instanton-induced (right) event-display in SHERPA; arrows represent particles produced in \pp\ collisions at $\sqrt{s}=13$ TeV. All hadrons with $p_{\rm T} > 0.2 $ GeV are shown.  }
    \label{fig:ParticleDistribution}
\end{figure}

The leading jet among dijet process is chosen such that its transverse momentum satisfies 
the criterion \pTjetL $> 25$~GeV, ensuring a well-defined hard scale in the perturbative regime. The 
sub-leading jet is selected with a lower threshold of $10$ GeV \ $<$ \pTjetSubL\ $<$ \pTjetL\ in order to remain 
sensitive to a semi-hard jet activity while avoiding an explicit bias toward symmetric dijet topologies. The subleading 
jet transverse momentum is required to satisfy \pTjetSubL\ $>$ 10 GeV/c, guided by measurements of the underlying event 
at the LHC~\cite{ALICE:2019mmy}, which demonstrate that soft particle production from multiple parton interactions 
dominates at low transverse momenta and saturates with the increasing hard scale. This requirement 
suppresses soft underlying event contributions while retaining sensitivity to semi-hard, isotropic multi-parton 
topologies characteristic of instanton-induced processes driven by the ’t Hooft vertex interaction~\cite{tHooft:1976rip}. Such an interaction leads to the production of several 
quarks and gluons with moderate transverse momenta in an approximately isotropic manner 
rather than a dominant back-to-back recoil structure. On the other hand, the chosen leading 
and sub-leading jet \JetPt\ ranges significantly suppress the pure underlying events. Figure~\ref{fig:ParticleDistribution} shows the particle production in a typical dijet and 
an instanton-induced event generated from \textsc{SHERPA}. All final state particles 
are presented in a two dimensional phase space consisting of the momenta along two spatial 
directions. Event topologies in these two cases are distinctly different. 
\subsection{Proposed new observables }

In this section, we propose two jet correlation observables that are sensitive probes of instanton-induced 
processes as compared to perturbative QCD dijet processes which are considered as the baseline. As discussed in 
section~\ref{sec:Jetreco}, instanton-induced events produce approximately isotropic final states. Dijet correlation (jet 
acoplanarity) observables are therefore natural discriminants as these probe whether a color-dipole structure exists at 
all, rather than how much it is smeared by soft-radiation. For the leading and sub-leading jets in an event, the 
acoplanarity angle is defined as $\Delta \phi = \phi_{\rm jet}^{\rm L} - \phi_{\rm jet}^{\rm SL}$ where $0 \le 
\Delta\phi \le \pi$. In a perturbative dijet production process, momentum conservation enforces a back-to-back topology 
with $\Delta\phi \simeq \pi$, while deviations arise from initial and final-state radiative processes. In contrast, 
instanton-induced processes do not originate from an underlying $2\to2$ scattering and therefore lack a preferred recoil 
axis, leading to a broad $\Delta\phi$ distribution. In this case, the number of associated sub-leading jets per leading 
jet is calculated as a function of \Delphi\,
\begin{equation}
    Y(\Delta \phi) = \frac{1}{N_{\rm jet}^{L}}\frac{dN_{\rm jet}^{\rm SL}}{d(\Delta\phi)}
\end{equation}
where $N_{\rm jet}^{L}$ and $N_{\rm jet}^{\rm SL}$ are the number of leading and sub-leading jets,
respectively. Similar measurements in a perturbative QCD context have been reported in both the LHC and RHIC 
experiments~\cite{CMS:2017cfb,ATLAS:2011kzm,STAR:2025yhg,ALICE:2023qve,ALICE:2023plt}.

To quantify the degree of azimuthal angular correlations among dijet events, the second harmonic moment
\begin{equation}
    \langle \cos(2\Delta\phi) \rangle =
\frac{1}{N_{\rm evt}} \sum_{\rm events} \cos(2\Delta\phi)
\end{equation}
of the particle distribution in $\Delta \phi$ is calculated. Here $N_{\rm evt}$ is the number 
of dijet events in a given \pTjetL\ bin. For perturbative dijet events, 
$\langle \cos(2\Delta\phi) \rangle \to 1$ as the transverse momentum of the jet increases, 
reflecting the recovery of a color-dipole topology. In contrast, for an approximately isotropic 
distribution of jet azimuthal angles, as expected for instanton-induced multi-parton final states, 
one obtains $    \langle \cos(n\Delta\phi) \rangle = 0 ~\forall ~ n \geq 1$.
The combined use of $Y(\Delta\phi)$ and $\langle \cos(2\Delta\phi) \rangle$ thus provides a sensitive 
discriminator between perturbative dijet and instanton-induced processes.

\subsection{Results and Discussions}
\label{Sec:resulstDiscussion}

\begin{figure}[!h]
    \centering
    \includegraphics[width=0.48\textwidth]{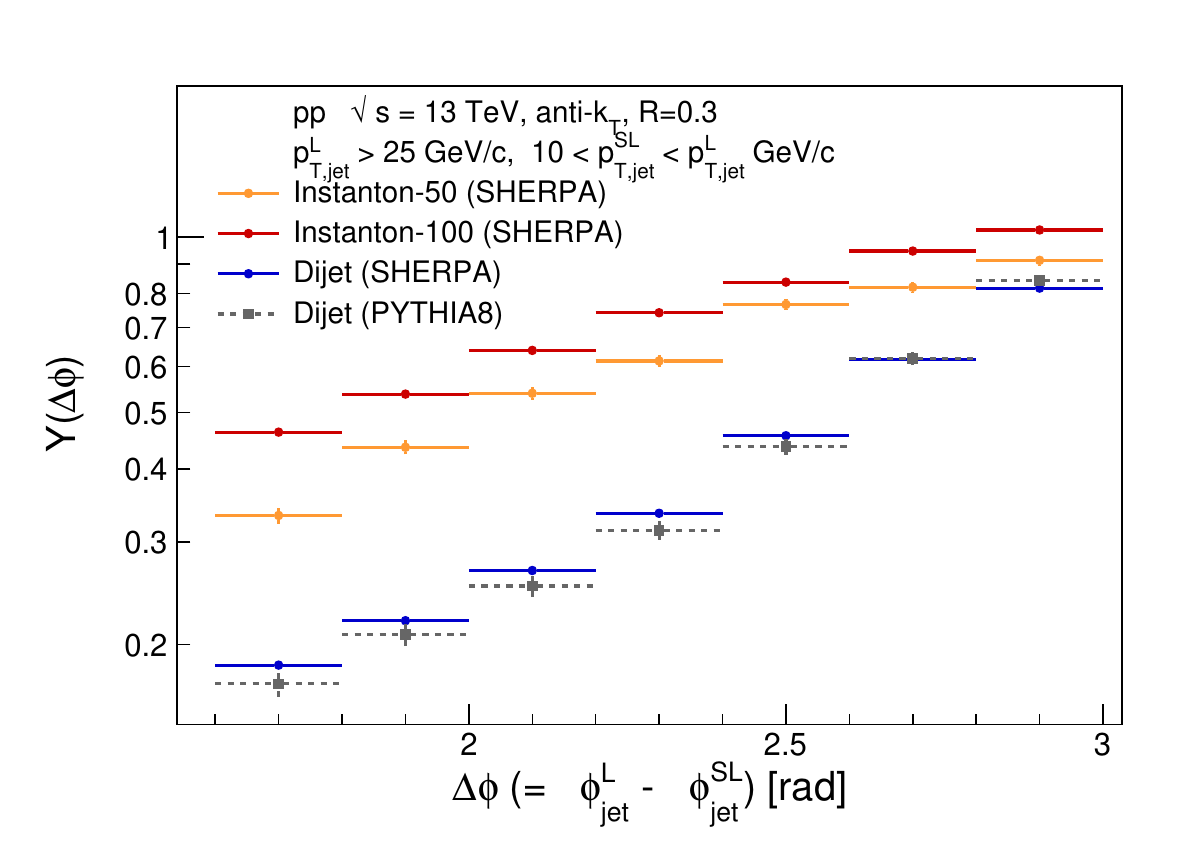}
    \caption{The \Delphi\ distributions corresponding to the total center-of-mass energy $50$ GeV (orange) and $100$ GeV (red) respectively, of the hadrons in the sub-leading jet and for dijet (blue ) events, both from SHERPA. Results for dijet events from PYTHIA8 are also shown for comparison as dotted lines.}
    \label{fig:Deltaphidistribution}
\end{figure}

The distribution of the dijet acoplanarity angle 
$\Delta\phi = \Big\vert\phi_{\rm jet}^{L} - \phi_{\rm jet}^{SL}\Big\vert$ for 
perturbative dijet events simulated with \textsc{PYTHIA8} and \textsc{SHERPA} is shown in 
Fig.~\ref{fig:Deltaphidistribution} and compared with the instanton-induced processes. For dijet correlations, the 
leading and sub-leading jets are chosen such that \pTjetL\ $>25$ GeV and $10$ GeV \ $<$ \pTjetSubL\ $<$ \pTjetL.  
The perturbative dijet samples exhibit a pronounced peak near $\Delta\phi \approx \pi$, reflecting the presence of an 
underlying back-to-back color-dipole topology, with a modest model dependence arising from different parton-shower and 
hadronization implementations.
\\

In contrast, instanton-induced events display a significantly broader $\Delta\phi$ distribution with a suppressed back-
to-back peak, consistent with the absence of a dominant $2\to2$ scattering axis and the production of multi-parton final 
states through the ’t~Hooft vertex interaction~\cite{tHooft:1976rip}. The observed separation between perturbative dijet 
and instanton topologies is substantially larger than the differences between \textsc{PYTHIA8} and \textsc{SHERPA}, 
demonstrating the discriminating power of the acoplanarity observable. While multi-jet and underlying-event 
contributions can also broaden angular correlations, they originate from perturbative radiative processes around 
a hard scattering axis, unlike instanton-induced processes which lack a primary $2 \rightarrow 2$ topology. 
In PYTHIA8 with MPI enabled, the semi-hard sub-leading jet selection ($10$ GeV $<$ \pTjetSubL $<$ \pTjetL ) 
suppresses soft underlying-event contributions, thereby reducing the impact of MPI on the jet correlation 
observables. A detailed study of MPI systematics and \pp\  high-multiplicity events is beyond the scope 
of this work.
 
\begin{figure}[!h]
    \centering
    \includegraphics[width=0.48\textwidth]{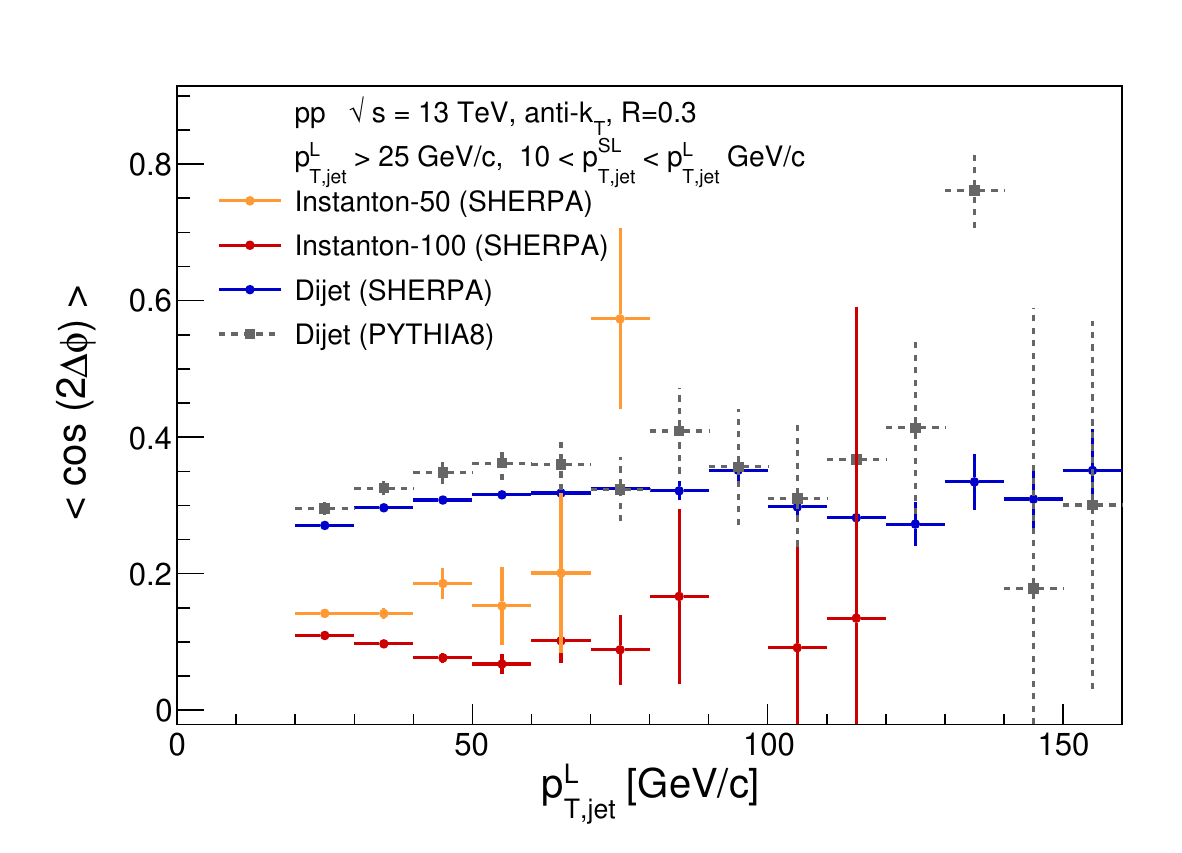}
        \caption{ \AveCosTwoDPhi\ vs. $p_{T,jet}^{L}$ for the same cases as discussed in Fig.~\ref{fig:Deltaphidistribution}.}
    \label{fig:cosTwoDeltaphi}
\end{figure}

The dependence of \AveCosTwoDPhi\ on the transverse momentum of the leading jet \pTjetL\  for perturbative dijet 
and instanton-induced events is shown in Figure~\ref{fig:cosTwoDeltaphi}. For dijets processes within 
\textsc{PYTHIA8} and \textsc{SHERPA}, the second harmonic increases with \pTjetL, reflecting the recovery of 
a back-to-back color-dipole topology at higher scales. Similar modest differences between the two models are 
observed here also. However, instanton-induced events exhibit a significantly suppressed \AveCosTwoDPhi\ over the full 
\pTjetL\ range, with no indication of recovery toward the dijet limit. Instanton-induced events thus characterized by a 
softer jet \pTjetL\ spectrum relative to perturbative dijet production consistent with their isotropic, multi-parton 
topology. The suppression is stronger for $\sqrt{s^{'}}=100$ GeV as compared to $50$ GeV consistent with isotropic 
multi-parton production picture due to larger contribution from small sized instantons.  The observed separation 
between perturbative dijet and instanton-induced topologies is significantly larger than the statistical fluctuations, 
indicating the robustness of the effect. Systematic variations associated with jet reconstruction and event selection, 
such as the choice of jet radius and transverse momentum thresholds, are expected to primarily affect the overall 
normalization but not the qualitative behavior of the acoplanarity observables.

The observed separation between perturbative dijet and instanton-induced topologies is consistent with the expectation 
that soft QCD radiation and MPI effects primarily broaden the underlying dipole structure, whereas instanton-induced 
processes lead to intrinsically more isotropic event topologies. To enhance sensitivity to instanton-induced topologies 
at the LHC, one may consider event selections based on high-multiplicity activity and semi-inclusive trigger jets with 
moderate to relatively lower energy thresholds. Such selections are expected to preferentially enhance isotropic, multi-
parton final states, while suppressing purely soft background contributions. A systematic investigation of these aspects 
is deferred to future studies.

\section{Summary}
\label{Sec:summary}

In this letter, we report for the first time, the results for the instanton size distribution and their relative 
separations in 2+1 flavor QCD with physical quark masses on a lattice with a finite cut-off. Though the results 
are consistent with the earlier reported trends in quenched QCD, however one naturally observes a suppression of the 
large sized instantons $\rho > 0.6$ fm which had to be earlier implemented in semiclassical calculations by including 
quantum corrections arising from both initial and final-state interactions in the instanton 
background~\cite{Khoze:2020tpp}. An understanding of the size and distance distribution allows us to explore  
in future the role of instantons in other processes in a hadron collider. Our lattice study thus allows to 
identify typical instanton sizes which can be described within instanton perturbation theory and for which 
it breaks down. Converting the $\sqrt{s^{'}}$ for the hadrons produced in an instanton-induced event simulated 
by event generators like SHERPA, to a size scale, and comparing with our lattice results allows us to show the 
reliability of semiclassical approximation that goes into calculating their cross-section. We use these event 
generators to generate perturbative dijet events in \pp\ collisions at \sqrtS\ $= 13$ TeV using representative 
values of the centre-of-mass energies of hadrons $\sqrt{s^{'}}=50$ and $100$ GeV respectively. We find that the 
instanton-induced events lead to a broad \Delphi\ distribution and its harmonic moment 
$\langle {\rm cos} (2\Delta \phi) \rangle$ exhibits a clear separation between a back-to-back perturbative 
dijet event and an instanton-induced event with a subleading jet. The proposed observables are experimentally 
accessible and provide a complementary approach to existing search strategies proposed for instanton-induced 
processes. \\

Several important directions merit further investigation. A systematic study of background contributions due to MPI, 
high-multiplicity QCD events, and jet reconstruction uncertainties, will be essential to quantify the sensitivity of 
these observables in realistic experimental conditions. In addition, exploring optimized event selections, such as 
semi-inclusive jet triggers and high-multiplicity criteria, could further enhance sensitivity to isotropic topologies. 
With the increased luminosity and improved detector capabilities in upcoming LHC runs, these measurements can be 
extended to larger datasets, enabling a more differential exploration of the acoplanarity observable as a function of 
jet kinematics.
\\

Furthermore, our approach is particularly promising for future measurements at the Electron–Ion Collider, 
where the kinematic variables $x$ and $Q^{2}$ provide direct control over the partonic structure and resolution scale 
of the interaction. The cleaner lepton–hadron environment, with reduced underlying events and multi-parton interactions, 
allows for a more direct probe of the absence of a $2\to 2$ scattering topology characteristic of instanton-induced processes. Combined measurements in \pp\ and \ep\ collisions will therefore offer a powerful strategy complementary to earlier HERA measurements to isolate topology-driven, non-perturbative effects in QCD.

\section*{Acknowledgments}
We are grateful to Rajiv V. Gavai, Peter Jacobs, Sven-Olaf
Moch and Edward Shuryak for discussions during the course of
this work. We acknowledge support from the Anusandhan National 
Research Foundation, Government of India, under Project ID
ANRF/ARG/2025/010416/PS. This research was supported in part by
the International Centre for Theoretical Sciences (ICTS) for participating
in the program International School and Workshop on Probing Hadron
Structure at the Electron-Ion Collider (ICTS/QEICIII2024/04). We also
acknowledge the use of high performance computing facility at the 
Institute of Mathematical Sciences.

\bibliographystyle{apsrev4-2}


\bibliography{Paper_Instanton}

@article{Bhattacharya:2014ara,
    author = "Bhattacharya, Tanmoy and others",
    title = "{QCD Phase Transition with Chiral Quarks and Physical Quark Masses}",
    eprint = "1402.5175",
    archivePrefix = "arXiv",
    primaryClass = "hep-lat",
    reportNumber = "BNL-103837-2014-JA, CU-TP-1205, INT-PUB-14-003, LLNL-JRNL-650194",
    doi = "10.1103/PhysRevLett.113.082001",
    journal = "Phys. Rev. Lett.",
    volume = "113",
    number = "8",
    pages = "082001",
    year = "2014"
}

@article{Khoze:2020tpp,
    author = "Khoze, Valentin V. and Milne, Daniel L. and Spannowsky, Michael",
    title = "{Searching for QCD Instantons at Hadron Colliders}",
    eprint = "2010.02287",
    archivePrefix = "arXiv",
    primaryClass = "hep-ph",
    reportNumber = "IPPP/20/44",
    doi = "10.1103/PhysRevD.103.014017",
    journal = "Phys. Rev. D",
    volume = "103",
    number = "1",
    pages = "014017",
    year = "2021"
}

@article{Amoroso:2020zrz,
    author = "Amoroso, Simone and Kar, Deepak and Schott, Matthias",
    title = "{How to discover QCD Instantons at the LHC}",
    eprint = "2012.09120",
    archivePrefix = "arXiv",
    primaryClass = "hep-ph",
    doi = "10.1140/epjc/s10052-021-09412-1",
    journal = "Eur. Phys. J. C",
    volume = "81",
    number = "7",
    pages = "624",
    year = "2021"
}

@article{Khoze:2021jkd,
    author = "Khoze, V. A. and Khoze, V. V. and Milne, D. L. and Ryskin, M. G.",
    title = "{Hunting for QCD instantons at the LHC in events with large rapidity gaps}",
    eprint = "2104.01861",
    archivePrefix = "arXiv",
    primaryClass = "hep-ph",
    reportNumber = "IPPP/20/93",
    doi = "10.1103/PhysRevD.104.054013",
    journal = "Phys. Rev. D",
    volume = "104",
    number = "5",
    pages = "054013",
    year = "2021"
}

@article{Aguilar:2021sfa,
    author = "Aguilar, Manny Rosales and Chang, Zilong and Elayavalli, Raghav Kunnawalkam and Fatemi, Renee and He, Yang and Ji, Yuanjing and Kalinkin, Dmitry and Kelsey, Matthew and Mooney, Isaac and Verkest, Veronica",
    title = "{pythia8 underlying event tune for RHIC energies}",
    eprint = "2110.09447",
    archivePrefix = "arXiv",
    primaryClass = "hep-ph",
    doi = "10.1103/PhysRevD.105.016011",
    journal = "Phys. Rev. D",
    volume = "105",
    number = "1",
    pages = "016011",
    year = "2022"
}

@article{Gleisberg:2008ta,
    author = "Gleisberg, T. and Hoeche, Stefan. and Krauss, F. and Schonherr, M. and Schumann, S. and Siegert, F. and Winter, J.",
    title = "{Event generation with SHERPA 1.1}",
    eprint = "0811.4622",
    archivePrefix = "arXiv",
    primaryClass = "hep-ph",
    reportNumber = "FERMILAB-PUB-08-477-T, SLAC-PUB-13420, ZU-TH-17-08, DCPT-08-138, IPPP-08-69, EDINBURGH-2008-30, MCNET-08-14",
    doi = "10.1088/1126-6708/2009/02/007",
    journal = "JHEP",
    volume = "02",
    pages = "007",
    year = "2009"
}

@article{Sherpa:2019gpd,
    author = "Bothmann, Enrico and others",
    collaboration = "Sherpa",
    title = "{Event Generation with Sherpa 2.2}",
    eprint = "1905.09127",
    archivePrefix = "arXiv",
    primaryClass = "hep-ph",
    reportNumber = "FERMILAB-PUB-19-218-T, SLAC-PUB-17433, IPPP/19/42, MCNET-19-11",
    doi = "10.21468/SciPostPhys.7.3.034",
    journal = "SciPost Phys.",
    volume = "7",
    number = "3",
    pages = "034",
    year = "2019"
}

@article{tHooft:1976rip,
    author = "'t Hooft, Gerard",
    editor = "Shifman, Mikhail A.",
    title = "{Symmetry Breaking Through Bell-Jackiw Anomalies}",
    reportNumber = "PRINT-76-0254 (HARVARD)",
    doi = "10.1103/PhysRevLett.37.8",
    journal = "Phys. Rev. Lett.",
    volume = "37",
    pages = "8--11",
    year = "1976"
}

@article{H1:2016jnv,
    author = "Andreev, Vladimir and others",
    collaboration = "H1",
    title = "{Search for QCD instanton-induced processes at HERA in the high- $\pmb {Q^2}$ domain}",
    eprint = "1603.05567",
    archivePrefix = "arXiv",
    primaryClass = "hep-ex",
    reportNumber = "DESY-16-050",
    doi = "10.1140/epjc/s10052-016-4194-6",
    journal = "Eur. Phys. J. C",
    volume = "76",
    number = "7",
    pages = "381",
    year = "2016"
}

@article{ZEUS:2003qkm,
    author = "Chekanov, S. and others",
    collaboration = "ZEUS",
    title = "{Search for QCD instanton induced events in deep inelastic ep scattering at HERA}",
    eprint = "hep-ex/0312048",
    archivePrefix = "arXiv",
    reportNumber = "DESY-03-201",
    doi = "10.1140/epjc/s2004-01735-3",
    journal = "Eur. Phys. J. C",
    volume = "34",
    pages = "255--265",
    year = "2004"
}

@article{CMS:2017cfb,
    author = "Sirunyan, Albert M and others",
    collaboration = "CMS",
    title = "{Azimuthal correlations for inclusive 2-jet, 3-jet, and 4-jet events in pp collisions at $\sqrt{s}= $ 13 TeV}",
    eprint = "1712.05471",
    archivePrefix = "arXiv",
    primaryClass = "hep-ex",
    reportNumber = "CMS-SMP-16-014, CERN-EP-2017-290",
    doi = "10.1140/epjc/s10052-018-6033-4",
    journal = "Eur. Phys. J. C",
    volume = "78",
    number = "7",
    pages = "566",
    year = "2018"
}

@article{ATLAS:2011kzm,
    author = "Aad, Georges and others",
    collaboration = "ATLAS",
    title = "{Measurement of Dijet Azimuthal Decorrelations in $pp$ Collisions at $\sqrt{s}=7$ TeV}",
    eprint = "1102.2696",
    archivePrefix = "arXiv",
    primaryClass = "hep-ex",
    doi = "10.1103/PhysRevLett.106.172002",
    journal = "Phys. Rev. Lett.",
    volume = "106",
    pages = "172002",
    year = "2011"
}

@article{STAR:2025yhg,
    author = "Aboona, B. E. and others",
    collaboration = "STAR",
    title = "{Measurement of medium-induced acoplanarity in central Au-Au and pp collisions at sNN=200 GeV using direct-photon+jet and {\ensuremath{\pi}}0 + jet correlations}",
    eprint = "2505.05789",
    archivePrefix = "arXiv",
    primaryClass = "nucl-ex",
    doi = "10.1103/k29c-d5ry",
    journal = "Phys. Rev. C",
    volume = "113",
    number = "1",
    pages = "014902",
    year = "2026"
}

@article{ALICE:2023qve,
    author = "Acharya, Shreyasi and others",
    collaboration = "ALICE",
    title = "{Observation of Medium-Induced Yield Enhancement and Acoplanarity Broadening of Low-pT Jets from Measurements in pp and Central Pb-Pb Collisions at sNN=5.02{\,}{\,}TeV}",
    eprint = "2308.16131",
    archivePrefix = "arXiv",
    primaryClass = "nucl-ex",
    reportNumber = "CERN-EP-2023-189",
    doi = "10.1103/PhysRevLett.133.022301",
    journal = "Phys. Rev. Lett.",
    volume = "133",
    number = "2",
    pages = "022301",
    year = "2024"
}

@article{H1:2002lra,
    author = "Adloff, C. and others",
    collaboration = "H1",
    title = "{Search for QCD instanton induced processes in deep inelastic scattering at HERA}",
    eprint = "hep-ex/0205078",
    archivePrefix = "arXiv",
    reportNumber = "DESY-02-062",
    doi = "10.1140/epjc/s2002-01039-8",
    journal = "Eur. Phys. J. C",
    volume = "25",
    pages = "495--509",
    year = "2002"
}

@inproceedings{Ringwald:1994kr,
    author = "Ringwald, A. and Schrempp, F.",
    title = "{Towards the phenomenology of QCD instanton induced particle production at HERA}",
    booktitle = "{8th International Seminar on High-energy Physics}",
    eprint = "hep-ph/9411217",
    archivePrefix = "arXiv",
    reportNumber = "DESY-94-197",
    month = "5",
    year = "1994"
}

@article{Moch:1996bs,
    author = "Moch, S. and Ringwald, A. and Schrempp, F.",
    title = "{Instantons in deep inelastic scattering: The Simplest process}",
    eprint = "hep-ph/9609445",
    archivePrefix = "arXiv",
    reportNumber = "DESY-96-202",
    doi = "10.1016/S0550-3213(97)00592-0",
    journal = "Nucl. Phys. B",
    volume = "507",
    pages = "134--156",
    year = "1997"
}

@article{Ringwald:1998ek,
    author = "Ringwald, A. and Schrempp, F.",
    title = "{Instanton induced cross-sections in deep inelastic scattering}",
    eprint = "hep-ph/9806528",
    archivePrefix = "arXiv",
    reportNumber = "DESY-98-081",
    doi = "10.1016/S0370-2693(98)00953-8",
    journal = "Phys. Lett. B",
    volume = "438",
    pages = "217--228",
    year = "1998"
}

@article{Ringwald:1999ze,
    author = "Ringwald, A. and Schrempp, F.",
    title = "{Confronting instanton perturbation theory with QCD lattice results}",
    eprint = "hep-lat/9903039",
    archivePrefix = "arXiv",
    reportNumber = "DESY-98-201",
    doi = "10.1016/S0370-2693(99)00682-6",
    journal = "Phys. Lett. B",
    volume = "459",
    pages = "249--258",
    year = "1999"
}

@article{Schafer:1996wv,
    author = {Sch{\"a}fer, Thomas and Shuryak, Edward V.},
    title = "{Instantons in QCD}",
    eprint = "hep-ph/9610451",
    archivePrefix = "arXiv",
    reportNumber = "DOE-ER-40561-293, INT-96-00-150",
    doi = "10.1103/RevModPhys.70.323",
    journal = "Rev. Mod. Phys.",
    volume = "70",
    pages = "323--426",
    year = "1998"
}

@article{Ringwald:1999jb,
    author = "Ringwald, A. and Schrempp, F.",
    title = "{QCDINS 2.0: A Monte Carlo generator for instanton induced processes in deep inelastic scattering}",
    eprint = "hep-ph/9911516",
    archivePrefix = "arXiv",
    reportNumber = "DESY-99-180",
    doi = "10.1016/S0010-4655(00)00148-X",
    journal = "Comput. Phys. Commun.",
    volume = "132",
    pages = "267--305",
    year = "2000"
}

@article{Balitsky:1986qn,
    author = "Balitsky, I. I. and Yung, A. V.",
    editor = "Shifman, Mikhail A.",
    title = "{Collective - Coordinate Method for Quasizero Modes}",
    doi = "10.1016/0370-2693(86)91471-1",
    journal = "Phys. Lett. B",
    volume = "168",
    pages = "113--119",
    year = "1986"
}

@article{Atiyah:1963zz,
    author = "Atiyah, M. F. and Singer, I. M.",
    title = "{The index of elliptic operators on compact manifolds}",
    doi = "10.1090/S0002-9904-1963-10957-X",
    journal = "Bull. Am. Math. Soc.",
    volume = "69",
    pages = "422--433",
    year = "1969"
}

@article{Hasenfratz:1998ri,
    author = "Hasenfratz, Peter and Laliena, Victor and Niedermayer, Ferenc",
    title = "{The Index theorem in QCD with a finite cutoff}",
    eprint = "hep-lat/9801021",
    archivePrefix = "arXiv",
    reportNumber = "BUTP-98-1",
    doi = "10.1016/S0370-2693(98)00315-3",
    journal = "Phys. Lett. B",
    volume = "427",
    pages = "125--131",
    year = "1998"
}

@article{Neuberger:1998my,
    author = "Neuberger, Herbert",
    title = "{A Practical implementation of the overlap Dirac operator}",
    eprint = "hep-lat/9806025",
    archivePrefix = "arXiv",
    reportNumber = "RU-98-28",
    doi = "10.1103/PhysRevLett.81.4060",
    journal = "Phys. Rev. Lett.",
    volume = "81",
    pages = "4060--4062",
    year = "1998"
}

@article{Narayanan:1994gw,
    author = "Narayanan, Rajamani and Neuberger, Herbert",
    title = "{A Construction of lattice chiral gauge theories}",
    eprint = "hep-th/9411108",
    archivePrefix = "arXiv",
    reportNumber = "IASSNS-HEP-94-99, RU-94-83",
    doi = "10.1016/0550-3213(95)00111-5",
    journal = "Nucl. Phys. B",
    volume = "443",
    pages = "305--385",
    year = "1995"
}

@article{Luscher:1998pqa,
    author = "Luscher, Martin",
    title = "{Exact chiral symmetry on the lattice and the Ginsparg-Wilson relation}",
    eprint = "hep-lat/9802011",
    archivePrefix = "arXiv",
    reportNumber = "DESY-98-014",
    doi = "10.1016/S0370-2693(98)00423-7",
    journal = "Phys. Lett. B",
    volume = "428",
    pages = "342--345",
    year = "1998"
}

@article{Kalkreuter:1995mm,
    author = "Kalkreuter, Thomas and Simma, Hubert",
    title = "{An Accelerated conjugate gradient algorithm to compute low lying eigenvalues: A Study for the Dirac operator in SU(2) lattice QCD}",
    eprint = "hep-lat/9507023",
    archivePrefix = "arXiv",
    reportNumber = "DESY-95-137, HUB-IEP-95-10",
    doi = "10.1016/0010-4655(95)00126-3",
    journal = "Comput. Phys. Commun.",
    volume = "93",
    pages = "33--47",
    year = "1996"
}

@article{Ginsparg:1981bj,
    author = "Ginsparg, Paul H. and Wilson, Kenneth G.",
    title = "{A Remnant of Chiral Symmetry on the Lattice}",
    reportNumber = "CLNS-81-520, HUTP-81-A060",
    doi = "10.1103/PhysRevD.25.2649",
    journal = "Phys. Rev. D",
    volume = "25",
    pages = "2649",
    year = "1982"
}

@article{Bonati:2015vqz,
    author = "Bonati, Claudio and D'Elia, Massimo and Mariti, Marco and Martinelli, Guido and Mesiti, Michele and Negro, Francesco and Sanfilippo, Francesco and Villadoro, Giovanni",
    title = "{Axion phenomenology and $\theta$-dependence from $N_f = 2+1$ lattice QCD}",
    eprint = "1512.06746",
    archivePrefix = "arXiv",
    primaryClass = "hep-lat",
    reportNumber = "IFUP-TH-2015-15",
    doi = "10.1007/JHEP03(2016)155",
    journal = "JHEP",
    volume = "03",
    pages = "155",
    year = "2016"
}

@article{Gavai:2024mcj,
    author = "Gavai, Rajiv V. and Jaensch, Mischa E. and Kaczmarek, Olaf and Karsch, Frithjof and Sarkar, Mugdha and Shanker, Ravi and Sharma, Sayantan and Sharma, Sipaz and Ueding, Tristan",
    title = {{Aspects of the chiral crossover transition in (2+1)-flavor QCD with M{\"o}bius domain-wall fermions}},
    eprint = "2411.10217",
    archivePrefix = "arXiv",
    primaryClass = "hep-lat",
    doi = "10.1103/PhysRevD.111.034507",
    journal = "Phys. Rev. D",
    volume = "111",
    number = "3",
    pages = "034507",
    year = "2025"
}

@article{Gross:2022hyw,
    author = "Gross, Franz and others",
    title = "{50 Years of Quantum Chromodynamics}",
    eprint = "2212.11107",
    archivePrefix = "arXiv",
    primaryClass = "hep-ph",
    doi = "10.1140/epjc/s10052-023-11949-2",
    journal = "Eur. Phys. J. C",
    volume = "83",
    pages = "1125",
    year = "2023"
}

@article{tHooft:1976snw,
    author = "'t Hooft, Gerard",
    editor = "Shifman, Mikhail A.",
    title = "{Computation of the Quantum Effects Due to a Four-Dimensional Pseudoparticle}",
    reportNumber = "PRINT-76-0551 (HARVARD)",
    doi = "10.1103/PhysRevD.14.3432",
    journal = "Phys. Rev. D",
    volume = "14",
    pages = "3432--3450",
    year = "1976",
    note = "[Erratum: Phys.Rev.D 18, 2199 (1978)]"
}

@article{Belavin:1975fg,
    author = "Belavin, A. A. and Polyakov, Alexander M. and Schwartz, A. S. and Tyupkin, Yu. S.",
    editor = "Taylor, J. C.",
    title = "{Pseudoparticle Solutions of the Yang-Mills Equations}",
    doi = "10.1016/0370-2693(75)90163-X",
    journal = "Phys. Lett. B",
    volume = "59",
    pages = "85--87",
    year = "1975"
}

@article{DeGrand:2000gq,
    author = "DeGrand, Thomas A. and Hasenfratz, Anna",
    title = "{Low lying fermion modes, topology and light hadrons in quenched QCD}",
    eprint = "hep-lat/0012021",
    archivePrefix = "arXiv",
    reportNumber = "COLO-HEP-452",
    doi = "10.1103/PhysRevD.64.034512",
    journal = "Phys. Rev. D",
    volume = "64",
    pages = "034512",
    year = "2001"
}

@article{Brower:2004xi,
    author = "Brower, Richard C. and Neff, Hartmut and Orginos, Kostas",
    editor = "Bodwin, Geoffrey T. and Sinclair, D. K. and Eichten, E. and Holmgren, D. and Kronfeld, Andreas S. and Mackenzie, P. and Okamoto, M. and Simone, J. and El-Khadra, Aida X.",
    title = "{Mobius fermions: Improved domain wall chiral fermions}",
    eprint = "hep-lat/0409118",
    archivePrefix = "arXiv",
    doi = "10.1016/j.nuclphysbps.2004.11.180",
    journal = "Nucl. Phys. B Proc. Suppl.",
    volume = "140",
    pages = "686--688",
    year = "2005"
}

@article{Bala:2025ilf,
    author = "Bala, Dibyendu and Kaczmarek, Olaf and Petreczky, Peter and Sharma, Sayantan and Tah, Swagatam",
    title = "{Spatial String Tension and Its Effects on Screening Correlators in a Thermal QCD Plasma}",
    eprint = "2501.17943",
    archivePrefix = "arXiv",
    primaryClass = "hep-lat",
    doi = "10.1103/3tmf-s94w",
    journal = "Phys. Rev. Lett.",
    volume = "135",
    number = "1",
    pages = "012301",
    year = "2025"
}

@article{Khoze:2019jta,
    author = "Khoze, Valentin V. and Krauss, Frank and Schott, Matthias",
    title = "{Large Effects from Small QCD Instantons: Making Soft Bombs at Hadron Colliders}",
    eprint = "1911.09726",
    archivePrefix = "arXiv",
    primaryClass = "hep-ph",
    reportNumber = "IPPP/19/85",
    doi = "10.1007/JHEP04(2020)201",
    journal = "JHEP",
    volume = "04",
    pages = "201",
    year = "2020"
}

@article{ALICE:2023plt,
    author = "Acharya, Shreyasi and others",
    collaboration = "ALICE",
    title = "{Search for jet quenching effects in high-multiplicity pp collisions at $ \sqrt{s} $ = 13 TeV via di-jet acoplanarity}",
    eprint = "2309.03788",
    archivePrefix = "arXiv",
    primaryClass = "hep-ex",
    reportNumber = "CERN-EP-2023-180",
    doi = "10.1007/JHEP05(2024)229",
    journal = "JHEP",
    volume = "05",
    pages = "229",
    year = "2024"
}

@article{ALICE:2019mmy,
    author = "Acharya, Shreyasi and others",
    collaboration = "ALICE",
    title = "{Underlying Event properties in pp collisions at $\sqrt{s}$ = 13 TeV}",
    eprint = "1910.14400",
    archivePrefix = "arXiv",
    primaryClass = "nucl-ex",
    reportNumber = "CERN-EP-2019-235",
    doi = "10.1007/JHEP04(2020)192",
    journal = "JHEP",
    volume = "04",
    pages = "192",
    year = "2020"
}
\end{document}